\documentclass[12pt]{iopart}
\usepackage{iopams}
\usepackage{graphicx}
\usepackage{dcolumn}
\usepackage{bm}
\usepackage{color}
\usepackage{latexsym}
\usepackage[T1]{fontenc}

\input{epsf}

\begin{document}


\date{\today}

\title{Anisotropy of strong pinning in multi-band superconductors}
\author{ C.J. van der Beek, M. Konczykowski}
\address{Laboratoire des Solides Irradi\'{e}s, CNRS UMR 7642 \& CEA-DSM-IRAMIS, Ecole Polytechnique, F-91128 Palaiseau cedex, France }
\author{R. Prozorov}
\address{The Ames Laboratory, Ames, IA 50011, U.S.A.}
\address{Department of Physics \& Astronomy, Iowa State University, Ames, IA 50011, U.S.A.}

\begin{abstract}
 The field-angular dependence and anisotropy of the critical current density in iron-based superconductors is evaluated using a phenomenological approach featuring distinct  anisotropy factors for the penetration depth and the coherence length. Both the weak collective pinning limit, and the strong pinning limit relevant for iron-based superconductors at low magnetic fields are considered.  It is found that in the more anisotropic materials, such as SmFeAsO and NdFeAsO, the field--angular dependence is completely dominated by the coherence--length (upper-critical field) anisotropy, thereby explaining recent results on the critical current in these materials. In less anisotropic superconductors, strong pinning can lead to an apparent inversion of the anisotropy. Finally, it is shown that, under all circumstances, the ratio of $c$--axis and $ab$--plane critical current densities for magnetic field along the $ab$--plane directly yields the coherence length anisotropy factor $\varepsilon_{\xi}$.
\end{abstract}


\maketitle

\section{Introduction}

The multi-band nature of superconductivity  in iron-based superconductors has now been well established \cite{Ding2008,Borisenko2010}. While direct measurement of the magnitude of the superconducting gap on the different Fermi-surface sheets has been and is the most convincing method in this respect, it is not the easiest. Therefore, many researchers have resorted to the measurement of the anisotropy of the superconducting parameters such as the coherence length and the penetration depth \cite{Welp2008,Welp2009,Pribulova2009,Kacmarcik2009,Okazaki2009,Song2010,Kim2011LiFeAsLambda} to prove the multi-band nature of superconductivity of these layered materials. This manifests itself through the inequality of the anisotropy factors $\varepsilon_{\xi} \equiv \xi_{c}/\xi_{ab}$ and $\varepsilon_{\lambda} \equiv \lambda_{ab}/\lambda_{c}$, which are the respective ratios of the coherence lengths $\xi_{c}$ and $\xi_{ab}$ parallel and perpendicular to the anisotropy ($c$)-axis, and of the penetration depths $\lambda_{ab}$ and $\lambda_{c}$ for supercurrents running in the $ab$--plane and along the $c$-axis respectively. The anisotropy of the coherence length is usually obtained from that of the upper critical fields $B_{c2}^{\perp}$ and $B_{c2}^{\parallel}$ for field oriented parallel to, and perpendicularly to the $c$-axis \cite{Welp2008,Welp2009,Pribulova2009,Kacmarcik2009,Song2010,Cho2011_LiFeAs_Hc2,Kurita2011Hc2}. The anisotropy of the penetration depth is most often extracted from the anisotropy of the lower critical fields $B_{c1}^{\perp} = (\Phi_{0}/4\pi\lambda_{ab}^{2}) \ln \kappa_{ab}$ and $B_{c1}^{\parallel} = (\Phi_{0}/4\pi\lambda_{ab}\lambda_{c}) \ln \kappa_{c}$ parallel and perpendicular to the $c-$axis \cite{Kacmarcik2009,Okazaki2009,Song2011EPL}, or from direct measurements using microwave techniques \cite{Hashimoto2008,Hashimoto2009,Gordon2009} (here $\kappa_{ab} \equiv \lambda_{ab}/\xi_{ab}$ while $\kappa_{c} \equiv \sqrt{\lambda_{ab}\lambda_{c}/\xi_{ab}\xi_{c}}$) \cite{Note_on_Torque}. The inequality of $\varepsilon_{\xi}$ and $\varepsilon_{\lambda}$ was first observed \cite{Hc2MgB2anisotropy,Lyard2004,Fletcher2005MgB2,Rydh2004}, and theoretically described \cite{Koshelev2003} in the archetypical two-band superconductor MgB$_{2}$. It was shown that in the case of  weakly coupled two-band superconductors such as MgB$_{2}$, Ginzburg-Landau theory, which, for single-band superconductors, introduces the anisotropy ratio $\varepsilon = \varepsilon_{\xi} = \varepsilon_{\lambda}$ as the square-root of the ratio of the effective masses $m_{ab}$ and  $m_{c}$ perpendicular and parallel to the $c$-axis, is inappropriate \cite{Koshelev2004}. 

\begin{figure}[t]
\includegraphics[width=1.1\linewidth]{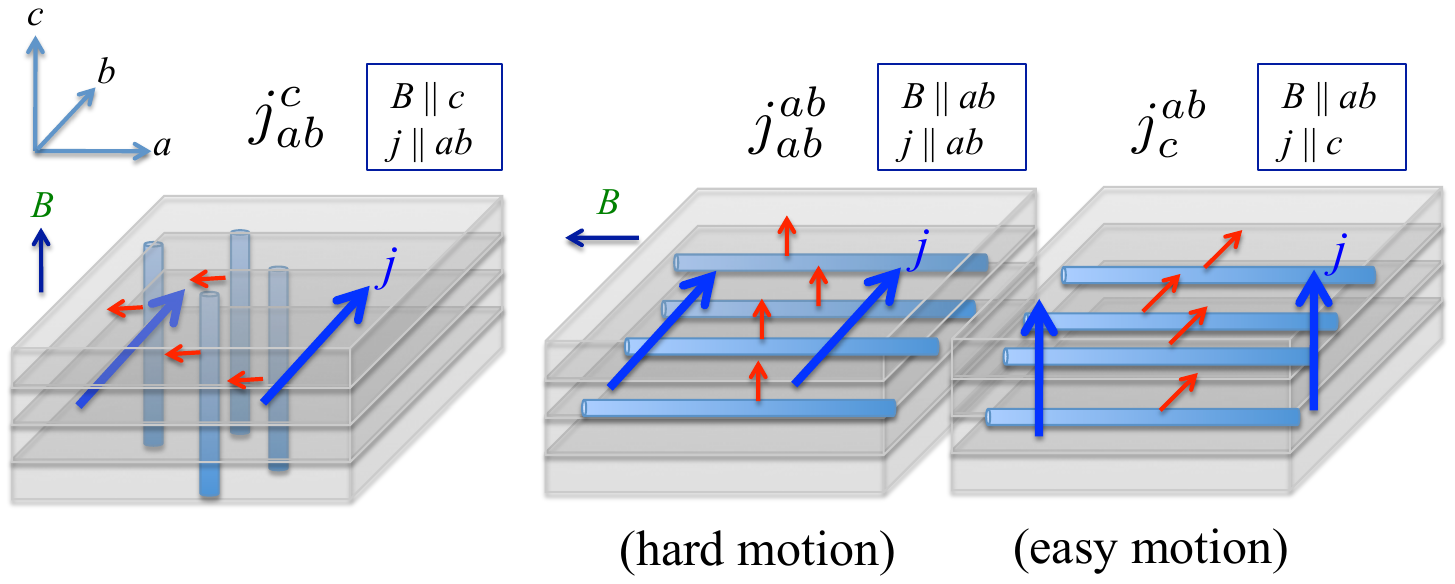}
\caption{Sketch defining the three independent critical current densities in an anisotropic superconductor. (a) $j_{ab}^{c}$ is the (most commonly measured) critical current density for current in the $ab$--plane, field along the $c$--axis. (b) the $ab$--plane critical current density $j_{ab}^{ab}$ for field in the $ab$--plane. This configuration involves the hard motion of vortex lines along the $c$-axis. (c) the $c$-axis critical current density $j_{c}^{ab}$ for field parallel to the $ab$--plane. This involves the easy motion of vortices along the $ab$--plane. }
\label{fig:anisotropy}
\end{figure}

More recently, critical current density measurements have been used to address the anisotropy of superconducting parameters in the iron-based superconductors. In particular, Kidzun {\em et al.}  \cite{Kidszun2011} and  H\"{a}nisch {\em et al.}  \cite{Hnisch} applied the anisotropic Ginzburg-Landau scaling procedure of Ref.~\cite{Blatter92}, developed for single band superconductors, to the critical current density of epitaxial LaFeAsO and Ba(Fe$_{1-x}$Co$_{x}$)$_{2}$As$_{2}$ films. They found  that the extracted temperature dependence of the  anisotropy parameter, $\varepsilon(T)$, matches that of the anisotropy parameter determined from scaling of the upper critical field \cite{Hnisch}. This appears to validate the use of critical current density measurements for obtaining reliable values of the anisotropy parameters. This approach was taken further by Moll  {\em et al.} \cite{Moll2010}, who specifically fashioned micrometer--sized SmFeAsO bridges to measure the three independent critical current densities, $j_{ab}^{c}$ for current running in the $ab$--plane and field oriented along the $c$-axis, and $j_{ab}^{ab}$ and $j_{c}^{ab}$ for current in the $ab$--plane and along the $c$--axis, respectively, and field in the $ab$-plane (see the definitions in Fig.~\ref{fig:anisotropy}). The three critical currents were also measured independently in single crystalline LiFeAs by using a Hall-probe array magnetometry technique \cite{Konczykowski2011}. These measurements allowed the direct extraction of the coherence length anisotropy.

However, the interpretation of the critical current anisotropy is, \em a priori \rm , not so straightforward. In principle, the critical current depends in a non-trivial manner on the specific nature of the pinning potential, that is, on the kind of defects and impurities that are present in the superconducting material.  It also depends non-trivially on the details of vortex elasticity. A supplementary complication comes from the fact that vortex pinning in the iron-based superconductors is, certainly at the low magnetic fields, dominated by the so-called strong pinning \cite{Ovchinnikov91,vdBeek2002}  by nm-sized impurities \cite{Kees,Kees1,Sultan}. In this regime, the anisotropy of the critical current may be influenced by the size and shape of the strongly pinning defects. Therefore, it is important to understand why the measurements of Refs.~\cite{Kidszun2011,Hnisch}, as well as those of Konczykowski {\em et al.} \cite{Konczykowski2011}, yield the coherence length anisotropy with such precision, even though the authors \cite{Kidszun2011,Hnisch} applied a single-band approach to analyze their results. 

In the following, we explore the effect of different anisotropies on the  critical current in the strong pinning regime in single-band and multi-band superconductors. Our goal will be limited to the identification of the main origin of the anisotropy of the critical current.  This justifies a purely phenomenological approach, in which we will suppose that, in a single-band superconductor, the angular dependence of the properties of the superconducting state enter through the angular dependence of the penetration depth, $\lambda(\vartheta) = \lambda_{ab} / \varepsilon_{\vartheta}$ for supercurrents running at an angle $\vartheta$ with respect to the $ab$--plane, while the angle--dependent coherence length $\xi(\vartheta) = \xi_{ab}\varepsilon_{\vartheta}$. Here $\vartheta$ is the angle between the orientation of interest and the $ab$--plane. The scaling function 
\begin{equation}
\varepsilon_{\vartheta} = \sqrt{\varepsilon^{2} \cos^{2} \vartheta + \sin^{2} \vartheta} < 1
\label{eq:scaling}
\end{equation}
can be found in Refs.~\cite{Blatter92,Blatter94}. For multi-band superconductors, we will assume that the angular dependence of vortex-related quantities in the mixed state can be described by two different anisotropy ratios $\varepsilon_{\lambda}(\vartheta)$ and $\varepsilon_{\xi}(\vartheta)$ for the penetration depth and the coherence length, and that the angular dependence of these ratios is the same as Eq.~(\ref{eq:scaling}). In all cases, $\varepsilon_{\xi}\xi$ will be assumed to be larger than the distance between the structural layers, so that the material can be described as ''continuously anisotropic''. Josephson tunneling between layers is beyond our consideration. 
 In section~\ref{subsection:weak} we first derive anew the results for the critical current density of Ref.~\cite{Blatter94}, for the case of weak collective pinning of individual vortices, keeping track of the anisotropies introduced by the coherence length and the penetration depth, respectively.  In section~\ref{section:strong}, we then apply the same procedure to the formulae of Ref.~\cite{vdBeek2002} for strong pinning by large point defects, which is more appropriate in iron-based superconductors at low fields. Here, the effect of a non-spherical shape of the pinning defects is also taken into account.

The procedure has the merit of demonstrating that the field-orientation--dependence of the $ab$--plane critical current density $j_{ab}(\vartheta)$ is, in all cases, determined by the competition of two factors. The coherence length anisotropy is responsible for the increase of the $j_{ab}(\vartheta)$ as the magnetic field is aligned towards the $ab$--plane, mainly because of the larger elementary pinning force experience by the vortex lines as these are turned towards the $ab$--plane. For large defects, this effect is counteracted by the penetration-depth anisotropy, which result in a larger line energy and therefore larger pinning energies when the vortices are aligned parallel to the $c$-axis. In most situations, the latter effect is much weaker than the first, so that the anisotropy extracted from field-angular dependent measurements of the critical current yields $\varepsilon_{\xi}$. 

The latter is also true for the anisotropy of the critical current density itself. On very general grounds, it is shown that the ratio $j_{c}^{ab}/j_{ab}^{ab}$ is, in all cases, equal to the coherence length anisotropy ratio $\varepsilon_{\xi}$. This fact was recently exploited to extract the field-dependence of $\varepsilon_{\xi}$ in single crystalline LiFeAs \cite{Konczykowski2011}.  We note  that the results below are relevant not only for iron-based superconductors, but also for high $T_{c}$ superconducting cuprate thin films, tapes, and composite conductors.

\section{Critical current anisotropy in uniaxial anisotropic superconductors}

\subsection{Weak collective pinning in single-band superconductors}
\label{section:single-band}
The anisotropy of the critical current density may, in principle, depend on the underlying pinning mechanism in the material. A flagrant example is the presence of anisotropic pinning centers such as heavy-ion irradiation-induced amorphous columnar defects \cite{Civale91,Nelson92,vdBeek95}, which introduce a  preferential orientation for the vortex lines supplementary to that imposed by the anisotropic material. However, in the case of weak collective pinning by dense atomic-scale point defects, the anisotropy of the critical current density ought not to depend on the details of the defect landscape, but only on the material properties. It can thus be straightforwardly obtained using the scaling formalism introduced by Blatter {\em et al.} \cite{Blatter92,Blatter94}. 

In the so-called single-vortex pinning regime, in which the critical current is not limited by intervortex interactions, the critical current density $j_{ab}$ in the ($ab$) plane, perpendicular to the symmetry ($c$)--axis, is expected to be field-angle independent, 
\begin{equation}
j_{ab}^{c}  =  \left(\frac{n_{d}\langle f_{p}^{2}\rangle \xi_{ab}^{2}}{\Phi_{0}L_{c}}\right)^{1/2}  = j_{ab}^{ab} =  \left( \frac{n_{d} \langle f_{p}^{\perp 2}\rangle \xi_{ab}\xi_{c}}{\Phi_{0}L_{c}^{ab}} \right)^{1/2}  \equiv j_{SV}.
\label{eq:jc-weak}
\end{equation}
\noindent Here, $f_{p}$ is the elementary pinning force of a single point defect for a vortex directed along the $c$-axis, $n_{d} \gg \xi^{-3}$ is the point defect density, and the factors $\xi_{ab}^{2}$ and $\xi_{ab}\xi_{c}$ derive from the statistical averaging of the pinning forces over the vortex core. The brackets $\langle \ldots \rangle$ denote averaging over the vortex core. For field aligned along $ab$, one is faced with a larger pinning force $f_{p}^{\perp}$ for ``out-of-plane'' vortex motion along $c$, since the pinning range \cite{Brandt86} $r_{f} \sim \xi_{c} = \varepsilon\xi_{ab}$ is shorter in this direction. However, this is  compensated by the larger longitudinal correlation length $L_{c}^{ab} = L_{c}/\varepsilon$ for field oriented along the $ab$-plane as compared to $L_{c}$ for field $\parallel c$ \cite{Blatter94}.  In this specific situation, $L_{c}^{ab}$  does not depend on the direction of the vortex displacement $\bf u$ in the pin potential. The anisotropic vortex line tensions, $\varepsilon_{1}^{\perp} = \varepsilon_{1}/\varepsilon_{}^{3}$ and $\varepsilon_{1}^{\parallel} = \varepsilon_{1}/\varepsilon_{}$ for out-of-plane and in-plane vortex displacements respectively, compensate for the anisotropies of the elementary pinning force and  the pinning range, \em i.e. \rm 
\begin{eqnarray}
L_{c}^{ab} & = & \left( \frac{ \varepsilon_{1}^{\perp}}{ n_{d}^{1/2} f_{p}/ \varepsilon} \right)^{2/3} \varepsilon^{1/3} \hspace{15mm} ({\bf u} \perp ab)
\label{eq:Lcab} \\
                    & =  & \left( \frac{ \varepsilon_{1}^{\parallel}}{ n_{d}^{1/2}  f_{p}} \right)^{2/3} \frac{1}{\varepsilon^{1/3}}  \hspace{18mm}({\bf u} \parallel ab) 
 \label{eq:Labab}\\
                    & = &  L_{c} /  \varepsilon .
                    \end{eqnarray}
Here $\varepsilon_{1} \sim \varepsilon^{2} \varepsilon_{0}$ and $\varepsilon_{0} = \Phi_{0}/4\pi\mu_{0}\lambda_{ab}^{2}$ are the vortex line tension and the vortex line energy for magnetic induction $\bf B \parallel c$, and $\Phi_{0}$ is the flux quantum. Hence, the critical current density $j_{ab}^{c}$ for field along the $c$-axis is equal to the critical current density $j_{ab}^{ab}$ (with $\bf B \perp \bf j$)  for field along the $ab$-plane.  

The $c$-axis critical current density $j_{c}^{ab}$ for  $\bf B \parallel ab$ on the other hand involves ``in-plane'' vortex motion, and is reduced by the anisotropy factor $\varepsilon$, such that  $j_{c}^{ab}=\varepsilon j_{ab}^{c}$. These relations do not hold in the so-called ``bundle regime'', in which vortex interactions reduce the magnitude of the critical current density. Then, $j_{ab}(B \varepsilon_{\vartheta})$ follows the well-known field-angular scaling relation  in $\varepsilon_{\vartheta} = (\varepsilon^{2} \cos^2 \vartheta + \sin^{2}\vartheta )^{1/2}$ as detailed in Ref.~\cite{Blatter94}. Here $\vartheta$ is the angle between the direction of the magnetic induction and the $ab$--plane. 

\subsection{Weak collective pinning in multi-band materials}
\label{subsection:weak}

We now turn our attention to the situation in the  multi-band iron-based superconductors. In what follows, the multi-band nature of superconductivity will be taken into account phenomenogically, by simply introducing different anisotropies $\varepsilon_{\xi}$ and $\varepsilon_{\lambda}$ of $\xi$ and $\lambda$, respectively. Then we can rederive the results or Ref.~\cite{Blatter94}, but keeping track of these different factors. We shall restrict ourselves to the single vortex regime of pinning, in which the critical current density is constant as function of magnetic field.  In iron-based superconductors, this is thought to be  relevant in a range of intermediate field strengths, sufficiently high for the critical current not be dominated by the strong pinning contribution due to nanometer-sized point-like pins (section~\ref{section:strong}), but below the onset of the so-called ''second magnetization peak''  regime, which is not described by the collective pinning theory \cite{Kees,Kees1}.  Typically, this corresponds to a field range of several tenths of Tesla, to a few Tesla at most.  

\subsubsection{Elementary pinning force}
The elementary pinning force $f_{p}$ can, within factors, be written as the product of the condensation energy $B_{c}^{2}/2\mu_{0} = \varepsilon_{0} / 2 \pi \xi_{ab}^{2}$, the defect volume $\frac{4}{3}\pi D_{v}^{3}$, and the inverse range of the pinning potential, $r_{f}^{-1}$. In the case of pinning by mean-free path variations \cite{Kees,Kees1,Blatter94,Thuneberg81}, a scattering factor proportional to $g(\rho_{D})(\xi_{0}/D_{v})$ should be introduced, with $g(\rho_D)$  the Gor'kov function,  $\rho_{D}= \hbar v_{F}/2\pi T_{c}l \sim \xi_{0}/l$  the disorder parameter,  $v_{F}$ the Fermi velocity, $l$ the mean free path, and $\xi_{0}  = 1.35\xi(0)$  the  temperature--independent  BCS coherence length. Disregarding the possibility of anisotropic scattering, the elementary pinning force for magnetic fields not aligned with the $c$-axis is modified because of the anisotropy of the vortex line energy $\varepsilon_{0}$, which involves the kinetic energy of current components flowing both parallel to $c$ and $ab$ as well as the anisotropy of the vortex core radius, and because of the anisotropic range of the pinning potential. When $\bf B$ is aligned along the $ab$--plane, the first effect introduces a factor $\varepsilon_{\lambda}$ and the second a factor $\varepsilon_{\xi}^{-1}$. As for the pinning range, this also depends on the  orientation of the supercurrent. For $\bf j \parallel ab$, $\bf j \perp \bf B$, vortex motion is parallel to the symmetry axis and, in the low--field  limit ($B < 0.2 B_{c2}^{\parallel}$), $r_{f}^{\perp} = \xi_{c}$, while for $\bf j \parallel c$,   $r_{f}^{\parallel} = \xi_{ab}$. Therefore,
\begin{eqnarray}
f_{p}^{\parallel} & = &  \frac{\varepsilon_{\lambda}}{\varepsilon_{\xi}} f_{p} \hspace{15mm}  ({\bf B} \parallel ab; {\bf j} \parallel c; |{\mathbf B}| < 0.2 B_{c2}^{\parallel}) 
\label{eq:fp-par}
\\
f_{p}^{\perp} & = &  \frac{\varepsilon_{\lambda}}{\varepsilon^{2}_{\xi}} f_{p}  \hspace{15mm} ({\bf B} \parallel ab; {\bf j} \parallel ab; {\bf j} \perp {\bf B}; |{\mathbf B}| < 0.2 B_{c2}^{\parallel}) .
\label{eq:fp-perp} 
\end{eqnarray}
For higher fields, the pinning range is given by the intervortex distance: $r_{f}^{\perp} = \varepsilon_{\xi}^{1/2}a_{0}$, while $r_{f}^{\parallel} = a_{0}/\varepsilon_{\xi}^{1/2}$ [with $a_{0} = (2 \Phi_{0}/\sqrt{3} B)^{1/2}$ the vortex spacing for ${\mathbf B} \parallel c$ ] \cite{Brandt86}. Therefore 
\begin{equation}
f_{p}^{\parallel}  =  \frac{ \varepsilon_{\lambda}}{\varepsilon_{\xi}^{1/2}}  f_{p}, \hspace{15mm}  ({\bf B} \parallel ab; {\bf j} \parallel c; |{\mathbf B}| > 0.2 B_{c2}^{\parallel}) 
\end{equation}
\noindent while
\begin{equation}
f_{p}^{\perp} =  \frac{\varepsilon_{\lambda}}{\varepsilon^{3/2}_{\xi}} f_{p}  \hspace{15mm} ({\bf B} \parallel ab; {\bf j} \parallel ab; {\bf j} \perp {\bf B}; |{\mathbf B}| > 0.2 B_{c2}^{\parallel}) .
\end{equation}

\subsubsection{Pinning correlation length}
The isotropy of the pinning correlation length for $\bf B \parallel ab$ in Eqs.~(\ref{eq:Lcab},\ref{eq:Labab}) is nominally lifted, because the anisotropy of the vortex line tension, $\varepsilon_{1}^{\perp} = \varepsilon_{1}/\varepsilon_{\lambda}^{3}$ and $\varepsilon_{1}^{\parallel} = \varepsilon_{1}/\varepsilon_{\lambda}$, bears mainly to that of $\lambda$, while the range $r_{f}$ is determined by the value of the coherence length in the direction of the vortex displacement. As a result, the pinning length $L_{c}^{\parallel} =  ( \varepsilon_{\xi}^{1/3} / \varepsilon_{\lambda}^{4/3} ) L_{c}$ of a vortex confined to the $ab$--plane will no longer be equal to $L_{c}^{\perp} = ( \varepsilon_{\xi}^{5/3} / \varepsilon^{8/3}_{\lambda} ) L_{c} = (\varepsilon_{\xi}/\varepsilon_{\lambda})^{4/2} L_{c}^{\parallel} $ for a vortex confined to a plane containing the $c$-axis. 

However, in a three-dimensional superconductor, both in--plane and out--of--plane random displacements are allowed, and both determine the metastable vortex position and the critical current density. Taking into account Eqs.~(\ref{eq:fp-par},\ref{eq:fp-perp}) and the difference of the line tensions for in-plane and out-of-plane vortex displacements, and minimizing the sum of elastic and pinning energy $\varepsilon_{1}^{\parallel} ( \xi_{ab}/L_{c}^{\parallel} )^{2}  + \varepsilon_{1}^{\perp} ( \xi_{c}/L_{c}^{\parallel} )^{2} - ( n_{d} \langle f_{p}^{\parallel 2} \rangle / L_{c}^{\parallel} )^{1/2} \xi_{ab}^{3/2}\xi_{c}^{1/2} - ( n_{d}\langle f_{p}^{\perp 2}\rangle / L_{c}^{\parallel} )^{1/2} \xi_{c}^{3/2}\xi_{ab}^{1/2} $, we obtain
\begin{equation}
L_{c}^{\parallel} = \frac{2^{2/3} \varepsilon_{\xi}^{1/3}}{\left( \varepsilon_{\lambda}^{2} + \varepsilon_{\xi}^{2} \right)^{2/3}} L_{c}.
\end{equation}
If the coherence length anisotropy dominates the penetration depth anisotropy, the vortex excursions will be mainly confined to the $ab$--plane,  the large pinning force in the $c$-direction prohibiting out-of-plane excursions. 

\subsubsection{Critical current density} As in Eq.~(\ref{eq:jc-weak}), the critical current density  is determined from the equality of the Lorentz force on a correlated pinned vortex segment and the total pinning force on that segment,
\begin{equation}
\Phi_{0} j_{c} L_{c} = \left( \frac{n_{d}\langle f_{p}^{2} \rangle}{L_{c}}\right)^{1/2} L_{c} \xi  \equiv F_{p}  
\label{eq:equality}
\end{equation}
where  $\xi$ is the square--root of the vortex core area. For the two different directions of motion of vortices parallel to the $ab$--plane, we find
\begin{eqnarray}
j_{c}^{ab} & = &  \frac{\varepsilon_{\lambda} \left( \varepsilon_{\lambda}^{2} + \varepsilon_{\xi}^{2} \right)^{1/3}}{2^{1/3} \varepsilon_{\xi}^{2/3}}  j_{SV} \hspace{10mm}  ({\bf B} \parallel ab; {\bf j} \parallel c) \\
j_{ab}^{ab} & = & \frac{\varepsilon_{\lambda} \left( \varepsilon_{\lambda}^{2} + \varepsilon_{\xi}^{2} \right)^{1/3}}{2^{1/3} \varepsilon_{\xi}^{5/3}}  j_{SV}   \hspace{8mm} ({\bf B} \parallel ab; {\bf j} \parallel ab; {\bf j} \perp {\bf B}).
\end{eqnarray}
Note that for $\varepsilon_{\lambda} \rightarrow \varepsilon_{\xi}$, $j_{c}^{ab} \rightarrow \varepsilon j_{SV}$ and $j_{ab}^{ab} \rightarrow j_{SV}$, in agreement with the results outlined in Section~\ref{section:single-band}. More importantly, the critical current density is independent of the range $r_{f}$ of the pin potential. Since the core area and the pinning length are the same, irrespective of the direction in which the vortices are driven, the difference between these expressions is entirely due to the different elementary pinning forces for in-plane and out--of--plane vortex motion. As a result, the anisotropy
\begin{equation}
\frac{j_{c}^{ab}}{j_{ab}^{ab}}  =  \varepsilon_{\xi}
\end{equation}
is uniquely determined by the anisotropy of the vortex core ({\em i.e.} of the coherence length). Therefore, the  measurement of the $j_{c}^{ab}/j_{ab}^{ab}$--anisotropy directly yields the coherence length ($B_{c2}$--) anisotropy. 

The above conclusion, a consequence of the fact that the critical current density in the single vortex regime can be written as $j_{c} = F_{p}/\Phi_{0}L_{c}$, can be extended to all field and temperature ranges. In particular, in the bundle pinning regime \cite{Blatter94}, Eq.~(\ref{eq:equality}) is replaced by 
\begin{equation}
j_{c}BV_{c} = \left( n_{d}\langle f_{p}^{2} \rangle V_{c} \right)^{1/2} 
\end{equation}
where the correlation volume $V_{c} = R_{c}^{2}L_{c}$ is the product of the square of the transverse correlation length $R_{c}$ and the longitudinal correlation length $L_{c}$. Since, for a given magnetic field orientation, the correlation volume is the same regardless of the direction in which the vortices are being driven, the anisotropy of the critical currents for different driving directions is always given by the anisotropy of the elementary pinning force. 

\subsection{Strong pinning in anisotropic superconductors}
\label{section:strong}
We now investigate the case of so--called strong pinning \cite{Ovchinnikov91,vdBeek2002,Blatter2004}, {\em i.e.}, the situation in which a type-II superconductor contains point-like defects or heterogeneities with typical dimensions $b_{z}$, $b_{x}$ parallel and perpendicular to the field direction larger the coherence length, and density $n_{i}$ much smaller than $\xi^{-3}$. Then, the (first moment of the) pinning force from different defects does not average out as in the weak pinning case. The critical current density is now determined by the direct sum of the forces exerted by each pin a vortex line is able to take advantage of \cite{Ovchinnikov91}. This situation is encountered in cuprate superconductor thin films \cite{vdBeek2002,Gutierrez2007} and coated conductors \cite{MacManusDriscoll2004}, in which second-phase nanoparticles are responsible for pinning, as well as in most iron-based superconductors at low fields \cite{Kees,Kees1,Sultan}.  In the latter materials, the presence of nm-scale heterogeneity of the superconducting properties is thought to play the role of the strong pins, thereby determining both the highly inhomogeneous vortex distributions, as well as the low-field critical current density \cite{Sultan}.

At low fields, the number of pins a vortex line can benefit from is solely determined by the vortex line tension. The mean characteristic length $\overline{\cal L} = \pi^{-1/2} \left( \varepsilon_{1}/n_{i}U_{p}\right)^{1/2}$ of free vortex between two effective pins is determined by the balance of the tilt deformation energy $\varepsilon_{1} ( u / {\cal L})^{2}$ of the line, and the pinning energy gain per pin, $U_{p}$. At magnetic inductions exceeding $B^{*}= \pi \Phi_{0} n_{i} \left( U_{p} /  \varepsilon_{0} \right)$, the intervortex repulsion rather than the line tension limits the number of defects any given flux line can take advantage of.  The pinning length is then given by $\overline{\cal L} = n_{i}u_{0}^{2}$, where the trapping area $u_{0}^{2} = U_{p} / (c_{66}\tilde{c}_{44})^{1/2} a_{0}$ is determined by the balance between the pinning energy gain and the energy of the lattice deformation. For ${\bf B} \parallel c$, the shear modulus is given by  $c_{66} \approx \varepsilon_{0}/4a_{0}^{2}$, and the non-local tilt modulus is evaluated as $\tilde{c}_{44} \approx \varepsilon^{2}\varepsilon_{0}/a_{0}^{2}$. For field along the $c$-axis, the critical current density is \cite{vdBeek2002} 
\begin{eqnarray}
j_{s} =  \frac{f}{\Phi_{0}\overline{\cal L}} & = & n_{i}^{1/2} \frac{f}{\Phi_{0}} \left( \frac{\pi U_{p}}{\varepsilon_{1}} \right)^{1/2}  \hspace{2cm} ( B < B^{*} ) 
\label{eq:jc-strong} \\
 & =  & n_{i}  \frac{f}{\Phi_{0}}  \frac{2 U_{p}}{ \varepsilon_{0}} \left( \frac{\Phi_{0}}{B}\right)^{1/2} \hspace{2cm} (B > B^{*})
\end{eqnarray}
where the maximum force that a single strong pin can exert on a vortex line (for ${\bf B} \parallel c$) 
\begin{equation}
f = \frac{B_{c}^{2}}{2 \mu_{0}} \frac{1}{r_{f}} \pi b_{z}\xi_{ab}^{2} \ln\left( 1 + \frac{b_{x}^{2}}{2\xi_{ab}^{2}} \right) 
  = \frac{1}{4} \varepsilon_{0}\frac{ b_{z}}{r_{f}} \ln\left( 1 + \frac{b_{x}^{2}}{2\xi_{ab}^{2}} \right)
  \label{eq:fp-strong}
\end{equation}
is the product of the condensation energy over the range of the pinning potential, the core volume occupied by the defect, and a logarithmic factor taking into account the modification of the supercurrent distribution around the defect \cite{vdBeek2002,Koshelev2011}.  Here we have assumed that the pinning defects are insulating, and of ellipsoidal shape with half-axes $b_{x} \parallel ab$ and $b_{z} \parallel c$. Eq.~(\ref{eq:fp-strong})  interpolates between the logarithmic dependence on dimensions expected for large defects with $b_{x},b_{z} \gg \xi_{ab}$, and the simple factor $(b_{x}/\xi_{ab})^{2}$ proportional to the defect cross-sectional area that one has for small defects.

\subsubsection{Elementary pinning force} 
In order to evaluate the anisotropy in the limit of strong pinning, both with respect to the field orientation and the driving direction, we again start by evaluating the change of the elementary force $f$ with field orientation.  Contrary to the case of weak pinning, $f$ is reduced by a factor $\varepsilon_{\lambda}$ for ${\mathbf B} \parallel ab$,  due to the decrease of  the vortex line energy. A supplementary complication  enters in that the smaller core cross-section for field not aligned with $c$, as well as the geometrical anisotropy of the defects, affect the factor under the logarithm. For ${\mathbf B} \parallel ab$ and ${\mathbf j} \parallel c$, {\em i.e.} in--plane vortex motion, the pinning range $r_{f} = \xi_{ab}$ is the same as for vortices  parallel to the $c$-axis, while for the motion along $c$ of vortices parallel to the $ab$, it is the smaller $c$-axis coherence length that enters. One thus has 
\begin{eqnarray}
f^{\parallel} & = &  \frac{\varepsilon_{\lambda}}{ \varepsilon_{b}(0)} f  \hspace{18mm}  ({\bf B} \parallel ab; {\bf j} \parallel c) \\
f^{\perp} & = &  \left(\frac{\varepsilon_{\lambda}}{\varepsilon_{\xi}\varepsilon_{b}(0)}\right) f;
\hspace{8mm} ({\bf B} \parallel ab; {\bf j} \parallel ab; {\bf j} \perp {\bf B})  
\end{eqnarray}
for arbitrary  orientation $\vartheta$ of the field with respect to the $c$-axis,
\begin{eqnarray}
f^{\parallel} & = & \frac{\varepsilon_{\lambda}(\vartheta)}{ \varepsilon_{b}(\vartheta)} f  \hspace{18mm}  ({\bf j} \parallel c) \\
f^{\perp} & = &  \left(\frac{\varepsilon_{\lambda}(\vartheta)}{\varepsilon_{b}(\vartheta)\varepsilon_{\xi}(\vartheta)}\right) f.
\hspace{8mm} ( {\bf j} \parallel ab; {\bf j} \perp {\bf B})  
\end{eqnarray}
where we have introduced the geometrical anisotropy
\begin{eqnarray}
\varepsilon_{b}(\vartheta) & = & \frac{1}{\sqrt{ \sin^{2}\vartheta + \frac{b_{x}^{2}}{b_{z}^{2}}\cos^{2}\vartheta }} \frac{\ln\left( 1 + b_{x}^{2}/2\xi_{ab}^{2}\right)}{ \ln \left[ 1 + b_{x}\sqrt{b_{x}^{2} \sin^{2}\vartheta + b_{z}^{2}\cos^{2}\vartheta}/2\xi_{ab}^{2}\varepsilon_{\xi}(\vartheta)  \right]} \nonumber \\
& & 
\end{eqnarray}
for model ellipsoidal defects. For spherical defects of size $\sim \xi_{ab}$, this reduces to a simple factor $\varepsilon_{\xi}(\vartheta)$. The pinning energy for ${\bf B}\parallel ab$ is $U_{p}^{ab} = [\varepsilon_{\lambda}/\varepsilon_{b}(0)]U_{p}$, while for arbitrary orientation it is $U_{p}(\vartheta) = [\varepsilon_{\lambda}(\vartheta)/\varepsilon_{b}(\vartheta)]U_{p}$. 

\subsubsection{Single vortex limit of strong pinning}
We first examine the limit of low fields. The pinning length $\overline{\cal L}$ for arbitrary field orientation takes on different values depending on the direction of the deformation: it is larger for out-of-plane deformations, due to the larger line tension $\varepsilon_{1}^{\perp} \sim \varepsilon_{1} / \varepsilon_{\lambda}^{3}(\vartheta)$ in this direction \cite{Blatter94}. At equilibrium, the most favourable pinned configuration will be determined by 
\begin{equation}
\overline{\cal L}^{\parallel} = \frac{\varepsilon_{b}^{1/2}(\vartheta)}{\varepsilon_{\lambda}(\vartheta)} \overline{\cal L}, \hspace{15mm}  ({\bf u} \parallel ab)
\end{equation}
since this is smaller than 
\begin{equation}
\overline{\cal L}^{\perp} = \frac{ \varepsilon_{b}^{1/2}(\vartheta)} {\varepsilon_{\lambda}^{2}(\vartheta)}  \overline{\cal L} \hspace{15mm}  ({\bf B} \parallel ab; {\bf u} \perp ab)
\end{equation}
by a factor $\varepsilon_{\lambda}(\vartheta) \equiv \sqrt{\varepsilon_{\lambda}^{2} \cos^{2} \vartheta + \sin^{2} \vartheta}$. Thus, the vortex adapts to the pin potential by  deformations parallel to the $ab$ plane rather than perpendicularly to the field direction. Note that in all cases, except that of extremely oblate defects ($\varepsilon_{b} < \varepsilon_{\lambda}^{2}$),  the typical length of a pinned segment $\parallel ab$ exceeds that for ${\mathbf B} \parallel c$, and thus, each vortex ought to be, on average, pinned by a smaller number of defects.  If one assumes that the vortices can seek out the most favorable pinned configuration at each stage of their motion, the $ab$--plane critical current density simply follows as
\begin{equation}
j_{ab}(\vartheta) =   \left( \frac{f^{\perp}}{\Phi_{0}\overline{\cal L}^{\parallel}} \right) = \frac{\varepsilon_{\lambda}^{2}(\vartheta)}{\varepsilon_{b}^{3/2}(\vartheta)\varepsilon_{\xi}(\vartheta)} j_{s};  \hspace{8mm} ({\bf j} \parallel ab; {\bf j} \perp {\bf B}; B < B^{*}_{\parallel} )  \nonumber \\
\hspace{4mm}  
\label{eq:single-strong}
\end{equation}
valid for $B < B^{*}_{\parallel}$ (see below). If the relevant pinning centers are relatively small (radius of the order of $\xi_{ab}$) and isotropic, the critical current density scales as $\varepsilon_{\lambda}^{2}(\vartheta)\varepsilon_{\xi}^{-5/2}(\vartheta)$. In the case of single-band superconductors, where $\varepsilon_{\lambda}(\vartheta) = \varepsilon_{\xi}(\vartheta) = \varepsilon_{\vartheta}$, this reduces to the scaling law $j_{ab}(\vartheta) \propto \varepsilon_{\vartheta}^{-1/2}$.

For ${\mathbf B} \parallel ab$, the critical current densities  parallel and perpendicular to the $c$-axis depend  on the anisotropies of the penetration depth and the coherence length in a different combination than in the weak pinning limit of section~(\ref{subsection:weak}):
\begin{eqnarray}
j_{c}^{ab} & = &  \left( \frac{f^{\parallel}}{\Phi_{0}\overline{\cal L}^{\parallel}} \right) = \frac{\varepsilon_{\lambda}^{2}(0)}{ \varepsilon_{b}^{3/2}(0)} j_{s} \hspace{18mm}  ({\bf j} \parallel c; B < B^{*}_{\parallel} ) 
\label{eq:parallel-low} \\
j_{ab}^{ab} & = &   \left( \frac{f^{\perp}}{\Phi_{0}\overline{\cal L}^{\parallel}} \right) = \frac{\varepsilon_{\lambda}^{2}(0)}{\varepsilon_{b}^{3/2}(\vartheta)\varepsilon_{\xi}(0)} j_{s};  \hspace{8mm} ({\bf j} \parallel ab; {\bf j} \perp {\bf B}; B < B^{*}_{\parallel} )  \nonumber \\
& & \hspace{40mm}  \label{eq:perp-low}
\end{eqnarray}
 However,  the generality of expression (\ref{eq:jc-strong}) together with a $\overline{\mathcal{L}}$ that does not depend on driving direction signifies that their ratio only depends on the coherence length anisotropy. In particular, for ${\mathbf B} \parallel ab$, the ratio of critical currents for in-plane and out-of-plane motion of vortices 
\begin{equation}
\frac{j_{c}^{ab}}{j_{ab}^{ab}}  =  \varepsilon_{\xi}
\end{equation}
again directly yields the coherence length anisotropy. 

\begin{figure}[t]%
\includegraphics[width=\textwidth]{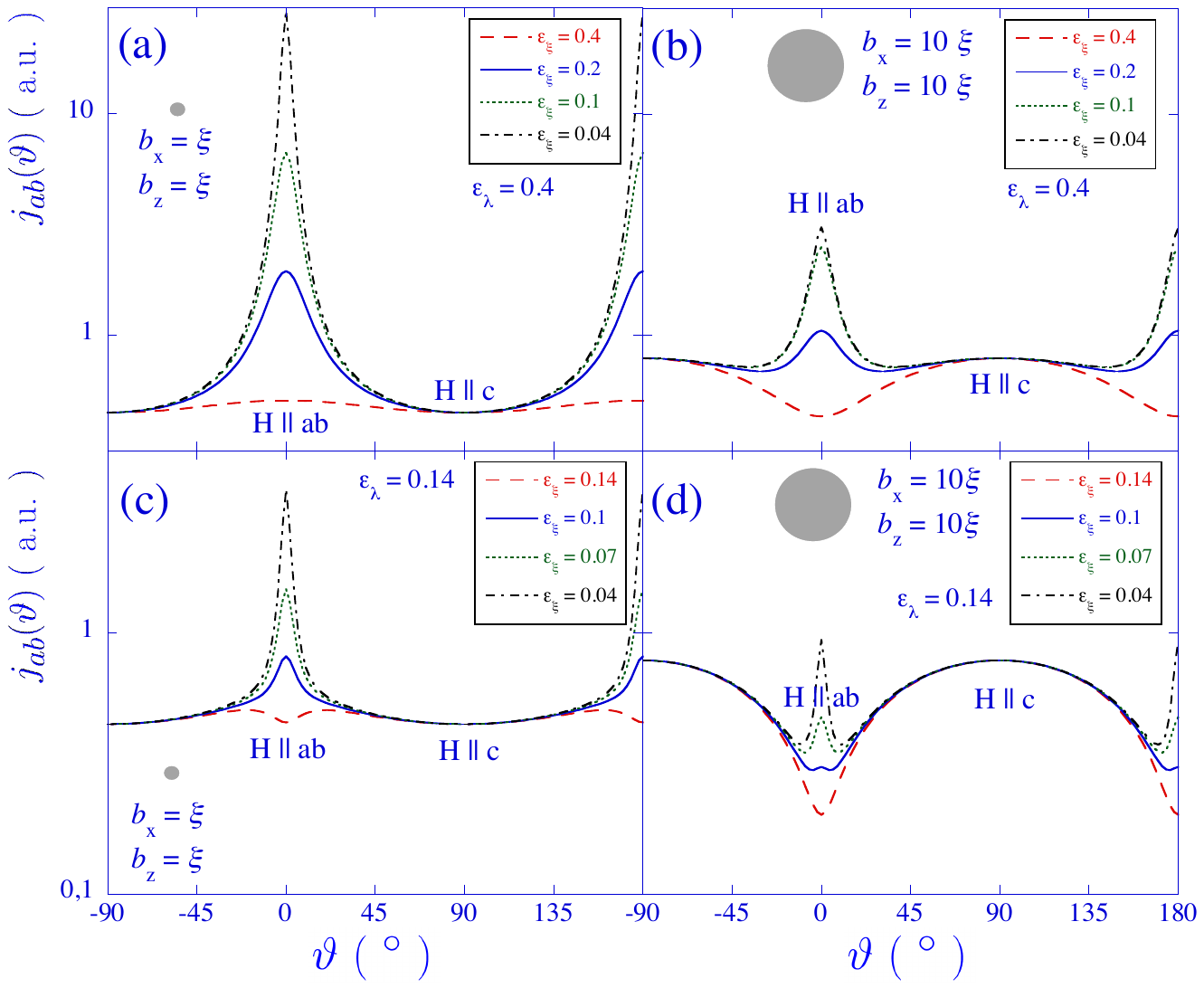}%
\caption{(Color online)  Angular dependence of the critical current density in the single vortex limit of strong pinning,  for anisotropies representative of (a,b) the (Ba,K)Fe$_{2}$As$_{2}$ system and (c,d) the NdFeAsO and SmFeAsO systems, containing spherical insulating defects. The different values  of $\varepsilon_{\xi}$ are introduced to mimic the temperature dependence  of the anisotropy. The graphs (a,c) depict the expected behaviour for small defect sizes, $b_{x} = b_{z} = \xi$, while panels (b,d) are for large defects with $b_{x} = b_{z} = 10 \xi$.  $\vartheta$ is the angle with respect to the $ab$--plane. }%
\label{fig-T-dependence}%
\end{figure}

\begin{figure}[h]%
\includegraphics[width=0.6\linewidth]{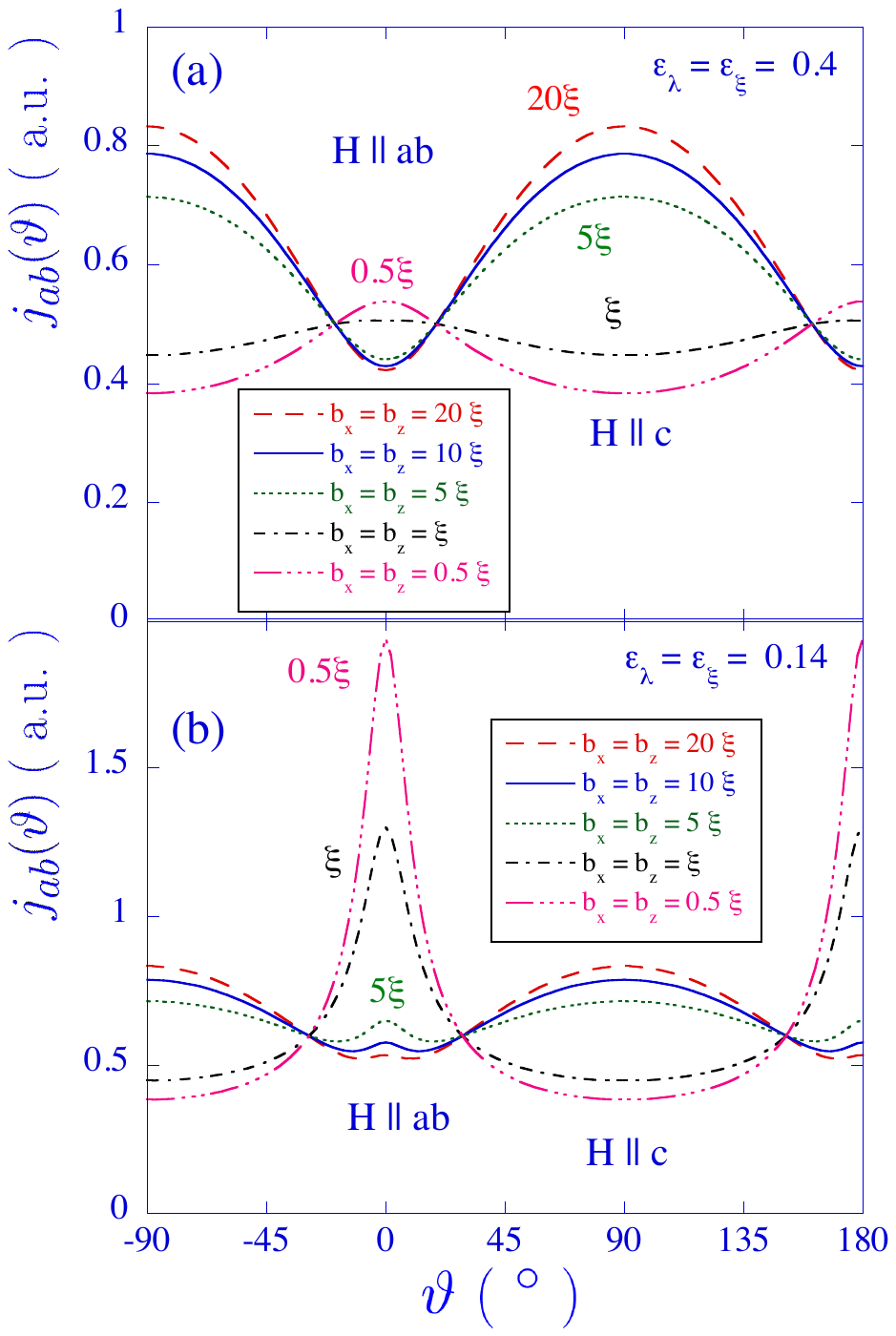}%
\caption{(Color online)   Angular dependence of the critical current density in the single vortex limit of strong pinning,  for anisotropies representative of (a) the (Ba,K)Fe$_{2}$As$_{2}$ system and (b) the NdFeAsO and SmFeAsO systems, for different radii $b_{x} = b_{z}$ of spherical insulating defects.  $\vartheta$ is the angle with respect to the $ab$--plane. }%
\label{fig-defect-size}%
\end{figure}

The case in which vortices cannot equilibrate during  driven motion was explored in Ref.~\cite{Koshelev2011}. Then, vortices can only take advantage of those pins that are located within a distance $u_{t}$ of the plane of motion.  The ``dynamic'' critical current density $j_{tr} = f/\Phi_{0}L$ for driven motion depends on the vortex length $L$ trapped between adjacent pins.  Analysis yields, for the critical current density for field along $c$, \cite{Koshelev2011} 
\begin{equation}
j_{tr} = n_{p}^{4/9} f  \left( \frac{U_{p}}{\varepsilon_{1}}\right)^{5/9} \xi^{-2/9}.
\end{equation}
The $ab$--plane critical current density for arbitrary field direction is obtained as using the parameters corresponding to (hard) motion, perpendicular to the field direction. Inserting relevant parameters, we obtain
\begin{eqnarray}
j_{ab,tr}(\vartheta) & = & n_{p}^{4/9} f^{\perp} \left[ \frac{U_{p}(\vartheta) \varepsilon_{\lambda}(\vartheta) }{\varepsilon_{b}(\vartheta)\varepsilon_{1}^{\perp}}\right]^{5/9}  \frac{1}{[\varepsilon_{\xi}(\vartheta)\xi ]^{2/9}} \\
& = & \frac{\varepsilon_{\lambda}^{29/9}(\vartheta)}{\varepsilon_{b}^{14/9}(\vartheta)\varepsilon_{\xi}^{11/9}(\vartheta)} j_{tr}.
\end{eqnarray} 
In the case of small spherical defects and similar anisotropies for the coherence length and the penetration depth, $j_{ab,tr}(\vartheta) \sim \varepsilon_{\vartheta}^{4/9}j_{tr}$. 

For field along the $ab$--plane, the critical current densities along $c$ and $ab$ are determined by hard and easy motion respectively, so that
\begin{eqnarray}
j_{c,tr}^{ab} & = & \frac{\varepsilon_{\lambda}^{19/9}}{\varepsilon_{b}^{14/9}(0)\varepsilon_{\xi}^{2/9}} j_{tr} \\
j_{ab,tr}^{ab} & = &  \frac{\varepsilon_{\lambda}^{29/9}}{\varepsilon_{b}^{14/9}(0)\varepsilon_{\xi}^{11/9}} j_{tr} .  
\end{eqnarray}
Unlike the above results, the ratio of the ``dynamic critical currents''  $j_{c,tr}^{ab}/j_{ab,tr}^{ab}  = \varepsilon_{\xi} / \varepsilon_{\lambda}^{10/9}$ does not uniquely depend on the coherence length anisotropy. Surprisingly, the derived dynamic critical current densities for vortex motion along the hard-- and the easy direction are nearly equal in single band materials, because the anisotropy of the elementary pinning force cancels  that of the vortex line tension.

\subsubsection{Three-dimensional strong pinning}
Turning to higher fields, our task is to determine the trapping radii $u_{0}^{\parallel}$ and $u_{0}^{\perp}$ 
for in-plane and out-of-plane vortex displacements, respectively. These are determined by the shear- and tilt moduli for vortex lattice deformations within the $ab$ plane, $c_{66}^{\parallel} = \varepsilon_{\lambda}^{3}(\vartheta) c_{66}$ and $c_{44}^{\parallel} \sim \tilde{c}_{44}$, and out-of-plane vortex deformations, $c_{66}^{\perp} = c_{66} / \varepsilon_{\lambda}(\vartheta)$ and $c_{44}^{\perp} \sim  \left( \varepsilon_{\lambda}(\theta)/\varepsilon_{\lambda}\right)^{2} \tilde{c}_{44}$ , respectively \cite{Sudbo91,Schonenberger93}.  Here $\theta = \frac{\pi}{2} -\vartheta$ is the angle with respect to the $c$-axis, and $\varepsilon_{\lambda}(\theta) \equiv \sqrt{\varepsilon_{\lambda}^{2} \sin^{2} \theta + \cos^{2} \theta}$ \cite{Blatter94}. Taking into account the distortion of the vortex lattice into account,  with in-plane and out-of-plane vortex spacings given by $a_{0}^{\parallel} = a_{0}/\varepsilon_{\lambda}^{1/2}(\vartheta)$, and $a_{0}^{\perp} = \varepsilon_{\lambda}^{1/2}(\vartheta) a_{0}$ respectively, we have 
\begin{eqnarray}
u_{0}^{\parallel 2}  & = & \frac{u_{0}^{2}}{  \varepsilon_{b}(0) }  \\
u_{0}^{\perp 2}  & = & \varepsilon_{\lambda} \left( \frac{\varepsilon_{\lambda}(\vartheta)}{\varepsilon_{b}(0)\varepsilon_{\lambda}(\theta)} \right) u_{0}^{2} ,
\end{eqnarray}
so that for ${\bf B} \parallel ab$,
\begin{eqnarray}
u_{0}^{\parallel 2}  & = &  \frac{u_{0}^{2}}{\varepsilon_{b}(0)}  \\
u_{0}^{\perp 2}  & = &\frac{\varepsilon_{\lambda}^{2}}{ \varepsilon_{b}(0)} u_{0}^{2} .
\end{eqnarray}
Thus, compared to the orientation $\parallel c$,  a vortex in the $ab$--plane can wander a similar distance in the plane, but a much smaller distance perpendicular to it. The trapping area $u_{0}^{\parallel}u_{0}^{\perp} = [\varepsilon_{\lambda}/\varepsilon_{b}(0)] u_{0}^{2}$, so that the length of a pinned vortex segment  $\overline{\cal L}^{\parallel} = 1/n_{i}u_{0}^{\parallel}u_{0}^{\perp}$ exceeds $\overline{\cal L}$ for ${\mathbf B} \parallel c$ by a factor $\varepsilon_{b}(0)/\varepsilon_{\lambda}$. The critical currents for vortex motion within the $ab$ plane and perpendicular to it are 
\begin{eqnarray}
j_{c}^{ab} & = &   \frac{f^{\parallel}}{\Phi_{0}}n_{i}u_{0}^{\parallel}u_{0}^{\perp}  = \frac{\varepsilon_{\lambda}^{2} }{\varepsilon_{b}^{2}(0)}  j_{s}^{c} \hspace{10mm}  ({\bf B} \parallel ab; {\bf j} \parallel c; B > B^{*}_{\parallel} ) 
\label{eq:parallel-high} \\
j_{ab}^{ab} & = &    \frac{f^{\perp}}{\Phi_{0}} n_{i}u_{0}^{\parallel}u_{0}^{\perp} = \frac{\varepsilon_{\lambda}^{2}}{\varepsilon_{b}^{2}(0)\varepsilon_{\xi}}  j_{s}^{c};  \hspace{8mm} ({\bf B} \parallel ab; {\bf j} \parallel ab. {\bf j} \perp {\bf B}; B > B^{*}_{\parallel})   
 \label{eq:perp-high} 
\end{eqnarray}
The crossover field $B^{*}_{\parallel} =  B^{*}/\varepsilon_{b}(0)$ delimiting the range of validity of Eqs.~(\ref{eq:parallel-high},\ref{eq:perp-high}) with respect to Eqs.~(\ref{eq:parallel-low},\ref{eq:perp-low}) is determined by comparing the high- and low field values of the respective critical current densities. It depends only on the specific details of the pinning centers.

 The ratio $j_{c}^{ab}/j_{ab}^{ab}  =  \varepsilon_{\xi}$ of the in-plane and out-of-plane critical current densities is once again determined by the coherence length anisotropy. Note that where, in the weak pinning regime, the critical current density for ${\bf B} \parallel ab$, ${\bf j} \parallel ab$, is the same as (or comparable to)  that  for  ${\bf B} \parallel c$, ${\bf j} \parallel ab$, in the strong pinning regime it may be noticeably smaller.

Finally, we give the expression for the field-dependent $ab$--plane critical current density as function of field orientation,
\begin{eqnarray}
j_{ab}(\vartheta) &  = & \frac{\varepsilon_{\lambda}^{2}(\vartheta)}{\varepsilon_{b}^{2}(\vartheta)\varepsilon_{\xi}(\vartheta)}  j_{s}^{c}\hspace{8mm} ( {\bf j} \parallel ab;  {\bf j} \perp {\bf B}; B > B^{*}_{\parallel})   
 \label{eq:angular-dependent-jc} \\
 & \propto &  j_{ab}^{ab}( B = 0 ) \frac{1}{[\varepsilon_{b}(\vartheta)B]^{1/2}} ,
\end{eqnarray}
valid for fields lower than  the angle-dependent crossover field $B^{*}(\vartheta) = B^{*}/\varepsilon_{b}(\vartheta)$. For small spherical defects in a single-band superconductor, this yields the well--known scaling law in $\varepsilon_{\vartheta}B$ \cite{Blatter92,Blatter94}.

\begin{figure}[h]%
\includegraphics[width=0.9\linewidth]{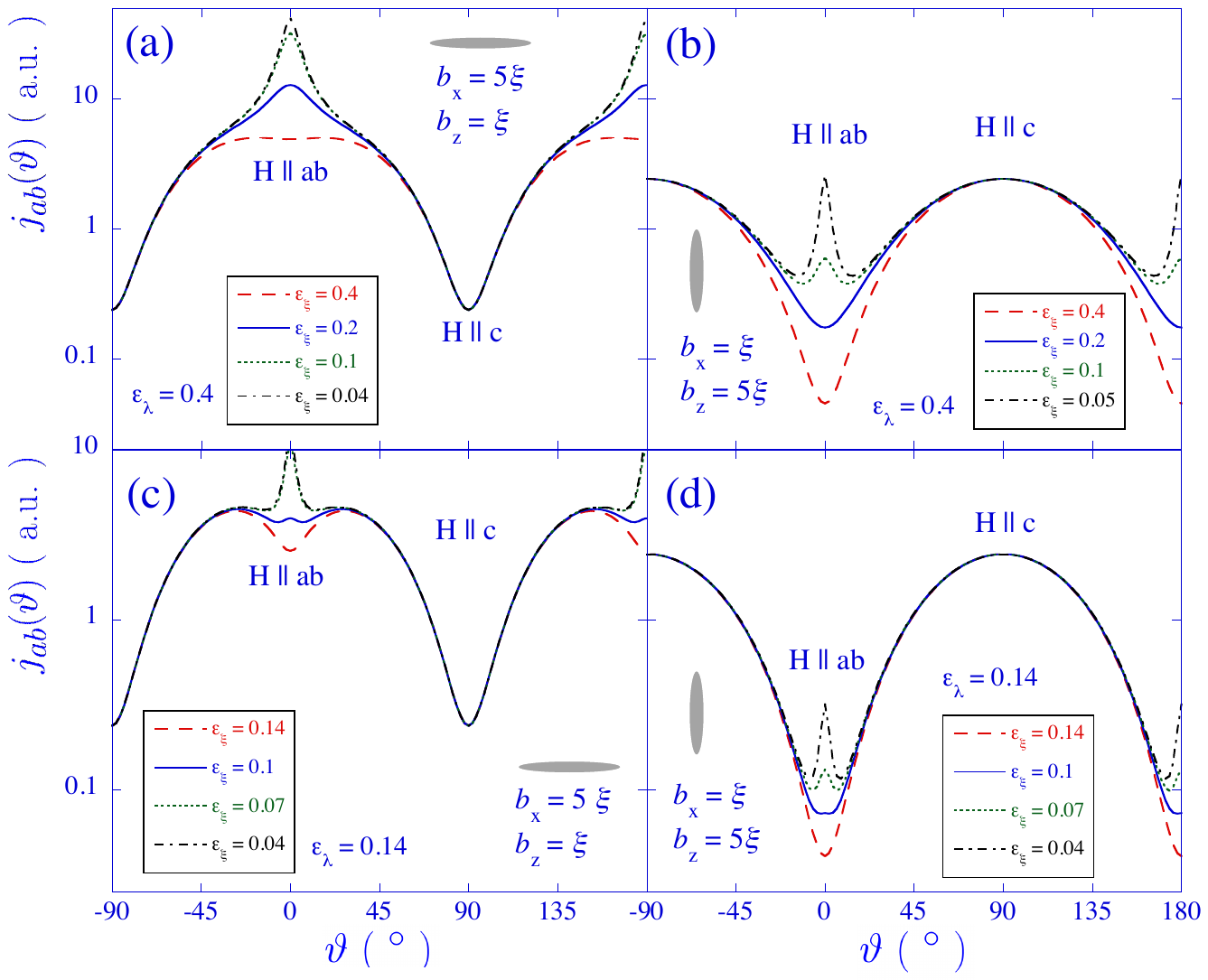}%
\caption{(Color online)  Angular dependence of the critical current density in the single vortex limit of strong pinning,  for anisotropies representative of (a,b) the (Ba,K)Fe$_{2}$As$_{2}$ system and (c,d) the NdFeAsO and SmFeAsO systems, for ellipsoidal oblate and prolate insulating defects with half axes  $b_{x} = 5\xi$, $ b_{z} = \xi$, and $b_{x} = \xi$, $b_{z} = 5 \xi$, respectively. $\vartheta$ is the angle with respect to the $ab$plane.  }%
\label{fig:oblate-prolate}%
\end{figure}

\section{Discussion}

The following salient features emerge from the above derivations. First is the field-angular dependence of the critical current density in the $ab$--plane. Eqs.~(\ref{eq:single-strong}) and (\ref{eq:angular-dependent-jc}) for the strong pinning critical current both involve the ratio of powers of the penetration depth anisotropy and the coherence length anisotropy. The penetration depth anisotropy in the denominator mainly comes from the line energy term appearing in the elementary pinning force. The coherence length anisotropy appears mainly due to the varying core size for different orientations, which determines the order parameter gradient and therefore, also, the magnitude of the pinning force. 

As a result, for small defects of size comparable to the coherence length, only the latter contribution matters. Fig.~\ref{fig-T-dependence}(a,c) shows the field-angular dependence of the $ab$--plane critical current density for such small defects, for two values of the penetration depth anisotropy, and different $\varepsilon_{\xi}$. Such a plot mimics the temperature dependence of pinning in the iron-based superconductors, in which $\varepsilon_{\xi} \lesssim \varepsilon_{\lambda}$ increases with temperature, the two reaching near equality near $T_{c}$. At low temperature $\varepsilon_{\xi} \ll \varepsilon_{\lambda}$ and the coherence length anisotropy completely dominates the behaviour. This is why thin films and crystals of the 1111-family of iron-based superconductors seem to behave as single-band anisotropic superconductors, and scaling of $j_{ab}$ with the product of the magnetic field and Eq.~(\ref{eq:scaling}) yields the $B_{c2}$--anisotropy \cite{Kidszun2011}. Note that the change of apparent anisotropy with changing $\varepsilon_{\xi}$  may be very difficult to observe; for example, as the temperature is increased, the absolute value of the coherence length also increases, so that defects that are to be considered ``large'' at low temperature become ``small'' close to $T_{c}$.

For larger defects, see Fig.~\ref{fig-T-dependence}(b,d), the scaling is disrupted, since now, the role of the angular-dependent vortex line energy becomes important. This results in a local maximum of the critical current for field aligned with the $c$-axis. For sufficiently large or strongly--pinning defects, this effect can result in the \em inversion \rm of the apparent material anisotropy. As summarized in Fig.~\ref{fig-defect-size}, this effect is not peculiar to multi-band superconductors, but is a property of strong flux pinning, as opposed to weak pinning.

The  effect of large defects on the field-angular dependence of the critical current density is particularly pronounced in the case of oblate and prolate defects, see Fig.~\ref{fig:oblate-prolate}. Note that this effect  is different in nature from that of extended columnar  \cite{Civale91,Nelson92,vdBeek95},  planar defects \cite{Kes89,Kwok90} or intrinsic pinning \cite{Ivlev90,Feinberg90,Kwok91,Balents94}, for which depinning takes place by half-- \cite{Nelson92,Ivlev90} or quarter--loop  \cite{Indenbom2000} nucleation rather than by pin-breaking. The nucleation--type depinning leads to a sharp maximum of the critical current density for field aligned with the defect direction, while the depinning from oriented ellipsoidal defects leads to a much broader maximum. In Fig.~\ref{fig:oblate-prolate}, the broad maxima of the critical current density as function of field angle correspond to the effect of the ellipsoidal defects, while the sharp structure, for field $\parallel ab$ is due to the material anisotropy. On top of that comes the possible effect of intrinsic pinning by the layered structure of the material, such as this may have been revealed by deviations from field angular-dependent scaling in Ref.~\cite{Kidszun2011}. The above analysis shows that in presence of strong pinning, the breakdown of simple scaling with $\varepsilon_\vartheta$ \cite{Kidszun2011} can arise because of several reasons, among which, the presence of point-like defects of dimensions larger than the coherence length, and the presence of defects of distinctly different dimensions and aspect ratios. 

A final and general conclusion is that the measurement of the critical current density anisotropy for field parallel to the $ab$--plane (perpendicular to the anisotropy axis of the material) is a useful tool for the direct determination of the coherence length anisotropy $\varepsilon_{\xi}$. It was shown above, on very general grounds, for both the weak-- and strong pinning scenarios, that the $j_{c}^{ab}/j_{ab}^{ab}$ ratio is always equal to $\varepsilon_{\xi}$. This is because the relative pinned volume, \em i.e. \rm the collective pinning length $L_{c}$ in the case of single vortex collective pinning, the correlation volume $V_{c}$ in the case of bundle pinning, and the pinned length $\overline{\mathcal L}$ in the case of strong pinning, only depends on the field orientation and not on the direction in which the vortices are being driven. Moreover, the critical current density does not depend on the range of the pinning potential. Therefore, only the anisotropy of the pinning force determines, in the end, the anisotropy of the critical current density. Since this is given by the anisotropy of the vortex core in a given field orientation, we may expect that it directly yields the coherence length anisotropy.

\begin{figure}[tb]%
\includegraphics[width=14cm]{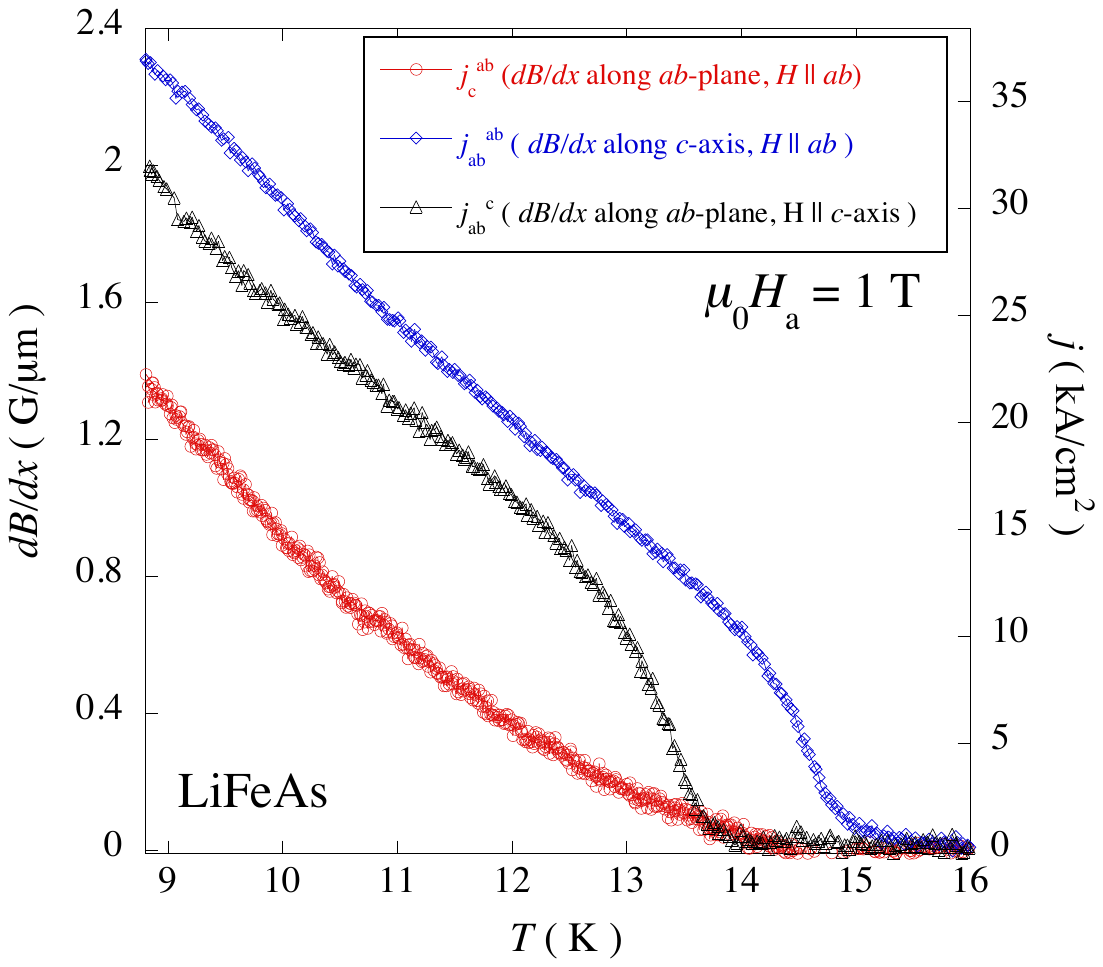}%
\caption{(Color online) Temperature dependence of the three independent critical current densities along the main crystalline axes of single crystalline LiFeAs, at an applied field of 1 T. The data depict $j_{c}^{ab}$ (\color{red}$\circ$\color{black}, current density along $c$, field along $ab$), $j_{ab}^{ab}$ (\color{blue}$\diamond$ \color{black}, current in the $ab$--plane, perpendicular to the magnetic field, also aligned in the $ab$--plane), and $j_{ab}^{c}$ ( $\triangle$, current in the $ab$--plane, field along the $c$--axis). The relevant magnitude of $j_{ab}^{ab}$ and  $j_{ab}^{c}$ suggest pinning either by insulating point defects of typical size $\sim \xi - 2 \xi$, or by areas larger than $2 \xi$ showing weaker superconductivity than the surrounding matrix.}%
\label{fig3}%
\end{figure}

\begin{figure}[t]%
\includegraphics[width=14cm]{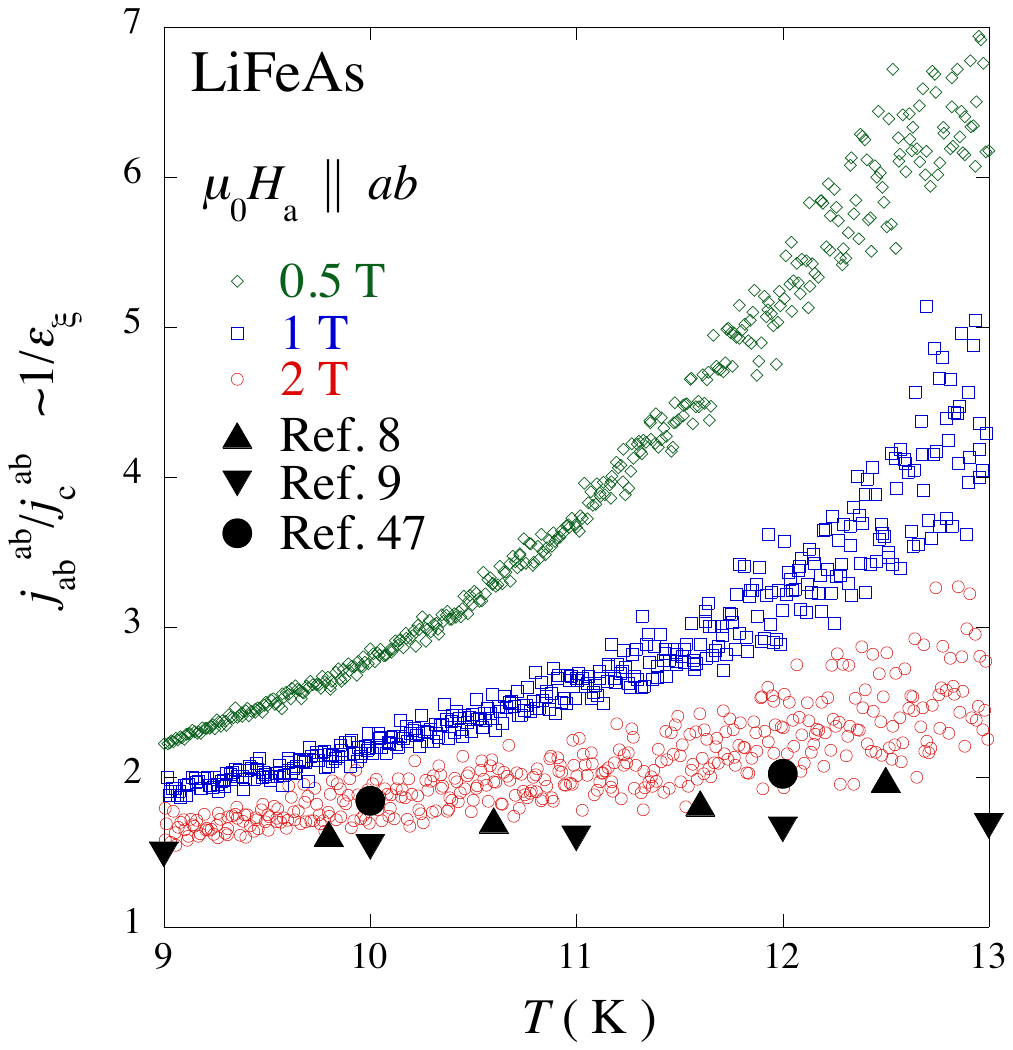}%
\caption{(Color online) Temperature dependence of the ratio $j_{ab}^{ab}/j_{c}^{ab}$, measured in single crystalline LiFeAs using a Hall probe array magnetometry technique \protect\cite{Konczykowski2011}, for applied magnetic fields of 0.5 T ( \color{green}$\diamond$ \color{black} ), 1 T (\color{blue}$\Box$\color{black}), and 2 T (\color{red}$\circ$\color{black}). Filled symbols show the anisotropy ratio $\varepsilon_{\xi}^{-1}$ determined from the anisotropy of the upper critical field  $B_{c2}$ in Refs.~\protect\cite{Cho2011_LiFeAs_Hc2}, \protect\cite{Kurita2011Hc2}, and \cite{Khim2011} . }%
\label{fig4}%
\end{figure}

The latter fact was recently exemplified by measurement of the three critical currents $j_{c}^{ab}$, $j_{ab}^{ab}$, and $j_{ab}^{c}$ on single crystalline LiFeAs \cite{Konczykowski2011}. Figure~\ref{fig3} shows the temperature dependence of the three critical currents, as determined by Hall array magnetometry of the flux density gradient on different crystal faces under different modes of field application.  From this, one may determine, directly, the ratio  $j_{c}^{ab}/j_{ab}^{ab}$ to yield the temperature-- and field--dependent coherence length anisotropy factor $\varepsilon_{\xi}(T,B)$. Fig.~\ref{fig4} shows that the result is in excellent agreement with the same quantity, determined from the upper critical field measurements of Refs.\cite{Cho2011_LiFeAs_Hc2,Kurita2011Hc2,Khim2011}.  Note that the anisotropy factor $\varepsilon_{\xi}$ has a non-trivial behaviour, in which it increases as function of field, but decreases as function of temperature. It can therefore not be simply explained in terms of the reduced field $B/B_{c2}$ at which the measurements were performed, in conjunction with  the temperature dependence $B_{c2}(T)$.

\section{Conclusion}

The field-angular dependence and anisotropy of the weak and strong  pinning critical current density in uniaxial anisotropic single-band and multi--band superconductors was considered, using a simple phenomenological approach introducing distinct anisotropy factors $\varepsilon_{\lambda}$ and $\varepsilon_{\xi}$ for the penetration depth and the coherence length respectively. Expressions for the field--angular dependent and anisotropic critical current density are derived. It turns out that the field-angular dependence of the $ab$--plane critical current density is determined by the competition between two effects. These are the angular variation of the vortex line energy, and therefore the pinning energy, which is maximum for field parallel to the $c$--(anisotropy) axis, and the angular variation of the vortex core size (coherence length), which yields larger elementary pinning forces for field parallel to $ab$. As a result, the field-angular dependence of the critical current density in materials with a relatively large coherence length anisotropy (small $\varepsilon_{\xi}$) and/or relatively small point--like pinning centers is indistinguishable from that in a single--band superconductor. Analysis of this angular dependence then yields the coherence length (upper critical field) anisotropy. On the other hand, in (single-band and multi-band) superconductors  with modest anisotropy and large point-like pinning centers, strong pinning for field oriented along the $c$-axis can lead to an apparent inversion of the anisotropy. Furthermore, we have shown, on very general grounds, that the ratio of the $c$-axis and $ab$--plane critical current densities for field along the $ab$--plane always yields the coherence length anisotropy $\varepsilon_{\xi}$. This can therefore be extracted as function of temperature and field, at magnetic inductions far below the upper critical field $B_{c2}$.

\section{Acknowledgments}
Work at Ames Laboratory was supported by the U.S. Department of Energy, Office of Basic Energy Sciences, Division of Materials Sciences and Engineering under contract No. DE-AC02-07CH11358. Work at SNU was supported by National Creative Research Initiative (2010-0018300). The work of R. Prozorov in Palaiseau was funded by the St. Gobain chair of the Ecole Polytechnique. C.J. van der Beek and M. Konczykowski acknowledge the hospitality of Ames Lab and Iowa State University during the preparation of this work.

\end{document}